# Efficient acousto-optical light modulation at the mid-infrared spectral range by planar semiconductor structures supporting guided modes

Ivan M. Sopko [1], Daria O. Ignatyeva [1,2,*], Grigory A. Knyazev [1,2], Vladimir I. Belotelov [1,2]

[1] Lomonosov Moscow State University, Faculty of Physics, Leninskie Gory, 119991, Moscow, Russia, [2] Russian Quantum Center, Novaya str., 143025, Skolkovo, Moscow, Russia

[*] ignatyeva@physics.msu.ru

**Acousto-optical devices, such as modulators, filters or deflectors, implement a simple and effective way of light modulation and signal processing techniques. However, their operation wavelengths are restricted to visible and near-infrared frequency region due to a quadratic decrease of the efficiency of acousto-optical interaction with the wavelength increase. At the same time, almost all materials with high value of acousto-optic figure of merit are non-transparent at wavelengths larger than 5 μm, while the transparent materials possess significantly lower acousto-optic figure of merit. Here we propose and demonstrate by calculations how these limitations could be overcome using specially designed planar semiconductor structures supporting electromagnetic modes strongly coupled to the incident light in the Otto configuration. Such approach could be used for a novel efficient acousto-optical device operating in middle infrared range of 8-14 μm. Acoustic wave excited by a piezoelectric transducer in a semi-conductor prism is utilized to modulate the coupling coefficient of the incident light to the semiconductor structure which results in up to 100% modulation of the transmitted light at the spatial scale less than the ultrasound wavelength. It allows to utilize acoustic waves with short decay distance and therefore, it provides a unique possibility to achieve an efficient acousto-optical modulation at frequencies over several gigahertz, which are unreachable for traditional acousto-optics.**

## Introduction

In this work we propose to utilize an acoustically controlled prism coupling of light in planar semiconductor structures supporting guided modes to enhance the acousto-optic interaction in the mid-infrared range, in particular at 10.6 μm wavelength. The importance of the considered topic is justified by the following reasoning. The vast majority of the acousto-optic devices is intended for visible and near-infrared frequency ranges [1], [2]. However, operation at the mid-infrared range is very promising due to atmosphere's transparency window at 8-14 μm wavelengths and the fact that thermal radiation maximum is within that range at room temperature. Nowadays there is an active development of mid-infrared tunable filters and image processing devices. In particular, there is a request for microscale light modulators with operational frequencies over 1 GHz [3]. In order to adapt acousto-optic devices and techniques to the mid and far-infrared ranges one has to overcome several challenges most notable of which is the lack of materials with high acousto-optic figure of merit that are transparent at wavelengths greater than 5-8 μm. Since the efficiency of the acousto-optic interaction decreases in the quadratic law with the wavelength growth the materials with extremely high acousto-optic figure of merit are necessary. Some of the unconventional materials proposed for such applications include single crystal tellurium [4-5], KRS-5 crystals [6], crystal iodic acid and lithium iodate [7]. Besides, the problem of operation at gigahertz frequencies is also quite challenging. The acoustic wave absorption scales quadratically with frequency which leads to the losses exceeding 10 dB/cm at 1 GHz for many acousto-optical materials [8]. As a result, the effective length of acousto-optical interaction is less than 100 μm which is negligible for bulk acousto-optics.

The significant enhancement of the acousto-optical interaction efficiency in mid-IR region could be realized via excitation of optical modes in a multilayer medium which are more sensitive than the bulk wave beams. Similar structures were successfully used for the enhancement of the light-matter interaction in layer structures [9], [10], where surface plasmon-polaritons [11], localized surface plasmons [12], waveguide modes [13], and Tamm plasmons are excited [14]. Acoustic modulation of the prism coupling coefficient with plasmonic structures was used to detect an acoustic wave by visible light in [15] and [16].

Plasmonics in the mid infrared range differs from conventional plasmonics of the visible range in terms of material properties and required techniques. The wavenumber of the surface-plasmon for most metals is very close to the wavenumber of light, so that surface plasmons poorly penetrate into the metal layer, while penetration depth in a dielectric significantly increases [17]. As a result, the differences between light incidence angle of the plasmon excitation and angle of the total internal reflection is negligible. This defines extremely narrow resonance shape [12], which is problematic for optimization of the device scheme but grants possibilities for high modulation depth [18]. Another approach is to utilize phonon-polaritons in a high absorbent medium like silicon carbide [19]. In that case phonon resonance is similar to the plasmon one in the visible range.

Here we consider two types of the structures: the positive permittivity semiconductor (PPS) structure consisting of a GaAs prism and thin film with an air gap in between, and the negative permittivity semiconductor (NPS) structure where a SiC film is utilized. The SiC film has negative permittivity near 10 μm due to the phonon resonance and supporting surface phonon-polariton mode. Modulation of the thickness of the gap between the prism and the semiconductor film sustaining optical mode results in a significant modulation of the coupling efficiency between the incident light and the guided mode, as well as in modulation of the mode propagation constant. It provides efficient modulation of the reflected light. Thus, we use Otto configuration for excitation of the guided modes. The advantage of the Otto configuration is in the fact that the gap thickness can be modulated via an acoustic wave propagating in the prism and reflecting from the gap boundary. As a result, there is no modulation of the guiding layer parameters. At the same time, because of high difference of acoustic impedances the depth of cladding modulation achieved by means of acoustic waves excited with piezoelectric transducer is estimated to be about 5 nm, which is approximately 0.05% of light wavelength. We show that this level of the acoustic modulation is enough to provide a large variation of the reflected light intensity up to 100% in the structures with high quality-factor optical resonances.

### Acousto-optical modulation based on coupling control

The principal scheme of the proposed modulator is shown in Fig.1 (a). The incident light impinges on the prism and due to the attenuated total internal reflection excites a waveguide mode in the PPS structure or a surface phonon-polariton in the NPS structure. We use Otto geometry of prism coupling since it allows for coupling control via acoustic waves as it is discussed below. A prism made from an infrared transparent material with high refractive index, such as GaAs or Ge is allocated above the multilayered structure with an air gap of about 1 μm between them. The longitudinal acoustic wave with the power density $W$, excited via piezoelectric transducer placed on top of the prism, provides variations of both the air gap thickness $\delta d$ and prism dielectric permittivity $\delta\varepsilon_p$ (see Supplementary).

The values of the air gap and dielectric permittivity modulation for different infrared materials are shown in Fig.1 (b). As follows from the data in Fig.1 (b) GaAs performance is superior to its competitors. Therefore, GaAs material was chosen as a material for the prism. One can notice that for materials such as $Bi_{12}GeO_{20}$, CdS and KRS-5 variation of dielectric permittivity $\delta\varepsilon$ is negative due to positive value of the photoelastic constant. Calculations show that in the case of positive photoelastic constants of $Bi_{12}GeO_{20}$, CdS and KRS-5 the influence of $\delta\varepsilon$ and $\delta d$ on the reflection coefficient is cumulative, and subtractive for materials with negative photoelastic constant such as GaP, Ge and GaAs. However, we chose to use GaAs prism for our research due to its relatively large absolute values of $\delta\varepsilon$ and $\delta d$. As it will be shown later, for the PPS structures, the impact of $\delta d$ variation is not significant and the response is determined almost fully by $\delta\varepsilon$, while for the NPS structures the situation is opposite. In this work calculations were performed for the acoustic frequency of 1GHz.

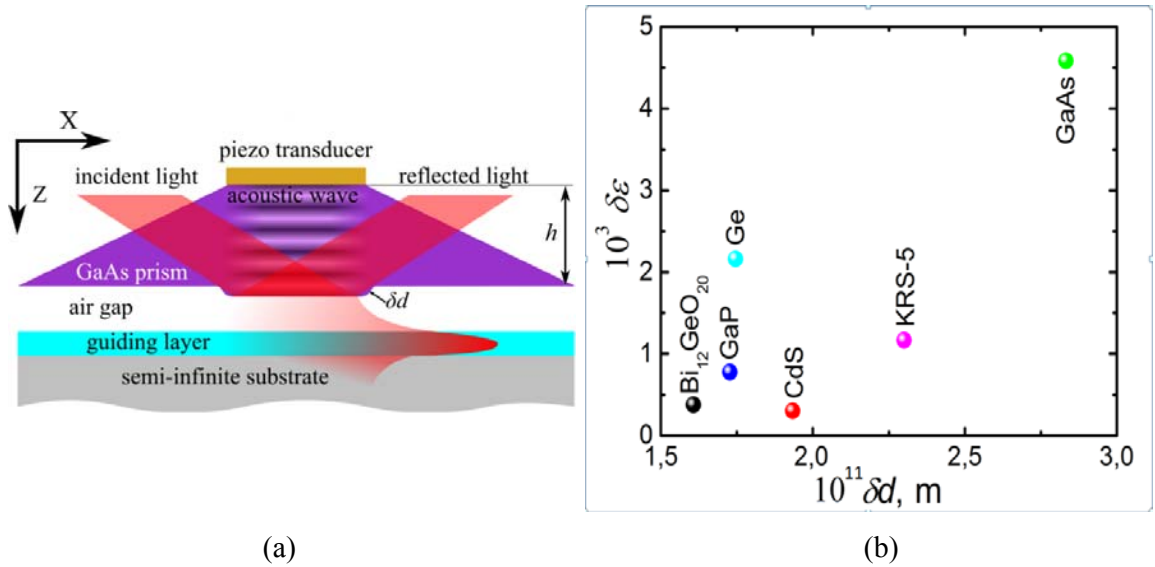

(a)                                                (b)

Fig.1 – Acousto-optical light modulation via excitation of optical modes in the Otto configuration: (a) modulator principle scheme; (b) comparison of different infrared materials for prism. The acoustic power density is assumed $W = 1 W/mm^2$, frequency of ultrasound is $f = 1 GHz$.

In order to analyze the influence of the prism dielectric permittivity and gap thickness variation on the coupling efficiency in Otto configuration, it is necessary to solve wave equations with boundary conditions for $E_y$ and $H_x$ for s-polarized light and for $E_x$ and $H_y$ for p-polarized light in the case of the attenuated total internal reflection at the prism-air interface. The method for calculation of reflection from the layered waveguide or surface wave supporting structures is mathematically similar to the one used to describe multilayered antireflection optical coating [20], and is generalized for the case of imaginary $k_z$ wave vector component arising in some layers of the analyzed structures.

## Acousto-optical modulation in the structure with negative permittivity semiconductor

It is advantageous to use silicon carbide for the NPS structure. It is a semiconductor material, which has a resonant absorbance peak at 10.6 μm with dielectric permittivity $\varepsilon_{SiC} = -1.46 + 0.15i$. That allows excitation of surface phonon-polaritons alongside the SiC-air interface. Just similarly to the surface plasmon-polaritons excitation of the surface phonon-polaritons is possible only in the case of p-polarized light. Phonons in IR range resemble plasmons at the visible range: it has a relatively wide shape and propagation length comparable to the wavelength of radiation. We start our study by analyzing surface phonon excitation on the silicon-carbide/air interface with the prism method in Otto configuration.

Implementing the methodology introduced in [14], [16], [19], [21] for the case of Otto configuration we get the reflection coefficient $R$ in the case of weak coupling, $\exp(-2k_{1z}d) \ll 1$:

$$R(k_x) = \left|\frac{k_{0z}/\varepsilon_p - k_{1z}/\varepsilon_{air}}{k_{0z}/\varepsilon_p + k_{1z}/\varepsilon_{air}}\right|^2 \left(1 - \frac{4\beta''\Delta\beta''}{[k_x - (\beta' + \Delta\beta')]^2 + (\beta'' + \Delta\beta'')^2}\right), \quad (1)$$

where $k_{0z}$ and $k_{1z}$ are wave vector component orthogonal to layers of the structure in the prism and in the airgap correspondingly, $\beta = \beta' + i\beta''$ is a wavenumber of the surface mode at the guiding-layer/air interface without consideration of the prism's influence. The prism coupling results in the wavenumber shift $\Delta\beta$. The real part of this shift, $\Delta\beta'$, corresponds to change in resonance position, while the imaginary part, $\Delta\beta''$, leads to the variations of the resonance width and depth symmetrically about the resonance center.

Obtaining expressions for the dispersion of the surface plasmon-polaritons on a thin film in asymmetric environment similarly to [20] one finds that the wavevector of the normal mode of the structure acquires additional term caused by the prism coupling:

$$\Delta\beta = \left(\frac{\omega}{c}\right) \frac{k_{0z}/\varepsilon_p - k_{1z}/\varepsilon_{air}}{k_{0z}/\varepsilon_p + k_{1z}/\varepsilon_{air}} \exp(-i2k_{1z}d)\, C(\varepsilon_j, d_j), \quad (2)$$

where $C$ is a parameter depending on the thickness and dielectric permittivities of the layers, not including the prism. The value of $C(\varepsilon_j, d_j)$ for the three-layer structure GaAs-air-SiC can be calculated in approximation $\varepsilon''_{SiC} \ll |\varepsilon'_{SiC}|$:

$$C = \frac{2}{\varepsilon_{SiC} - \varepsilon_{air}} \left(\frac{\varepsilon_{SiC}\varepsilon_{air}}{\varepsilon_{SiC}^2 + \varepsilon_{air}^2}\right)^{3/2}, \quad (3)$$

It should be noted, that the influence of air gap thickness variation $\delta d$ and change in dielectric permittivity of the prism $\delta\varepsilon$ do not depend on the structure properties. Besides, due to total internal reflection $k_{1z}$ is imaginary, so the exponent of power in (2) is real. Thus, the acoustic wave affects both resonance shape and position, however reflectance is modulated mostly by the resonance shift.

It's worth noting, that the addition to the phonon-polariton wavenumber is equivalent to the variation of the light incidence angle, at which the resonance is observed: $\Delta\beta = \Delta\theta \cos\theta \sqrt{\varepsilon_p}\,\omega/c$. As the acoustic interaction occurs through the prism coupling by means of $\delta d$ and $\delta\varepsilon$ variations, the resonance shift $\delta\theta$ caused by acoustic wave can be acquired from Eq. (2):

$$\delta\theta|_{\varepsilon=const} = -2\delta d \frac{|k_{1z}|}{k_{0z}} (\Delta\beta' + i\Delta\beta''), \quad (4)$$

$$\delta\theta|_{d=const} = \delta\varepsilon \frac{|k_{1z}|(\varepsilon_{air} + \varepsilon_p)}{k_{0z}^2 \varepsilon_{air}^2 - k_{1z}^2 \varepsilon_p^2} (\Delta\beta'' - i\Delta\beta'). \quad (5)$$

It is noteworthy, that the value of radiation losses can also be controlled by the air gap thickness. As follows from Eq.(1), the optimal value of $d$ is determined by the condition $\Delta\beta'' = \beta''$. As the angular shift $\delta\theta$ caused by $\delta\varepsilon$ variation depends on $\Delta\beta''$ linearly, the contribution of $\delta\varepsilon$ to the modulation of reflectance is weakly dependent on the resonance width, while the contribution of $\delta d$ is strongly affected by the resonance width. Further analysis shows, that at the given value of the acoustic power density the variation of the prism's dielectric permittivity $\delta\varepsilon$ corresponds to the shift angle $\delta\theta = 0.1°$ for examined structures. It means that for resonances narrower than 1° the contribution of $\delta\varepsilon$ prevails over $\delta d$ influence, while for wider resonances $\delta d$ is most important.

In order to calculate the modulation efficiency caused by acoustic waves we calculate reflection indexes for two full shift positions: $R_1=R(d+\delta d, \varepsilon_p-\delta\varepsilon_p)$ and $R_2=R(d-\delta d, \varepsilon_p+\delta\varepsilon_p)$. The modulation of the reflection coefficient $\zeta$ is defined by the formula:

$$\zeta = \frac{\delta R}{2<R>} = \frac{R(d-\delta d,\varepsilon_p-\delta\varepsilon_p)-R(d+\delta d,\varepsilon_p+\delta\varepsilon_p)}{R(d-\delta d,\varepsilon_p-\delta\varepsilon_p)+R(d+\delta d,\varepsilon_p+\delta\varepsilon_p)}. \tag{6}$$

In the basic case of GaAs-prism/air-gap/semi-infinite-SiC-layer the deepest resonance of 18% reflectance is achieved for the air gap thickness of 970 nm at incidence angle value of 45.1°. In Fig 2(a) the reflectance and the modulation coefficient $\zeta$ for the air gap thickness modulation of 0.03 nm are represented. In the inset a temporal dependence of reflectance in case of harmonic acoustic oscillation is shown. It is seen that modulation is close to sinusoidal shape. This trend is explained by the low level of modulation coefficient $\zeta < 1.3\%$.

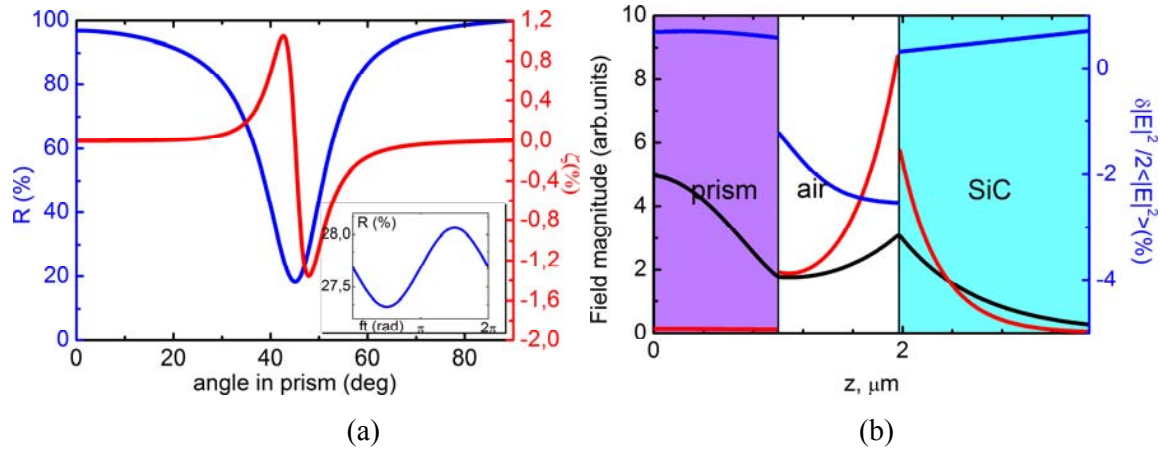

Fig.2 Acousto-optical modulation enhanced via excitation of phonon-polaritons in SiC. (a) Angular spectra of the reflectance R and modulation coefficient $\zeta$ and temporal dependence of $R$ (inset). (b) Spatial distribution of the electromagnetic field $H_y$ (black) and $|\mathbf{E}|^2$ (red) and its modulation $\delta|\mathbf{E}|^2/2<|\mathbf{E}|^2>$ (blue) due to acoustic oscillations inside the structure.

The resonance is rather wide, $\Delta\theta=15°$, therefore, $\delta d$ plays much greater role than $\delta\varepsilon$. If we assume that the acoustic wave causes $\delta\varepsilon = 3\cdot10^{-3}$ and air gap variation of $\delta d = 2.83$ nm (see Fig.1b), the contribution of the later would be 3.5 times greater. This configuration due to wide resonance shape posses very high tolerance of operating wavelength variation – less than 0.0011 %/nm.

Fig.2b demonstrates a spatial distribution of electromagnetic field. The distribution of field proves that on the surface of SiC a surface phonon-polariton is excited. As shown by the blue curve the modulation of the electric field caused by $\delta d$ achieves the highest values on the surface of SiC but not in the prism. It means that three layer NPS structure is not optimal for modulation of light. This curve is plotted for the incidence angle corresponded to the highest modulation of $R$ (see Fig.1a).

In order to achieve deeper and narrower resonances it is possible to utilize thin SiC films on the substrate (for example, ZnS with $n=2.21$) that allows for simultaneous control of coupling coefficient and internal damping. Best results were achieved for the NPS structure containing 570 nm thick SiC layer with 433 nm air gap. The resulting resonance reaches $R=0.05\%$ at 52.7° (see Fig.3a). This minimum reflectance value was chosen to mimic small background always present due to roughness of the layer surfaces and width inaccuracies, in order to avoid artificial overstatement of modulation efficiency due to near-zero denominator in Eq. (8). The angle of total internal reflection for GaAs/ZnS interface is 42.52° which results in a singular spike on the graph. This structure provides almost 20% modulation. Small non-linear distortions of modulation of $R$ are observed in the inset of Fig.3a. The described scheme, unlike the one with semi-infinite SiC crystal, is more sensitive to operating wavelength variation – less than 2 %/nm. It is even more sensible to variation of the SiC crystal thickness – about 30 %/nm.

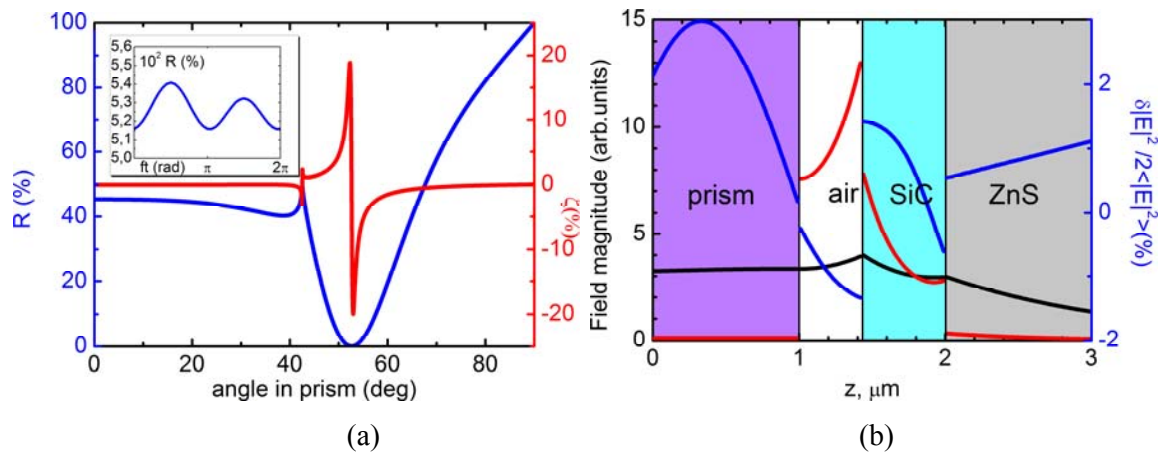

(a)   (b)

Fig.3 Acousto-optical modulation enhanced via excitation of phonon-polaritons in the SiC layer with ZnS substrate. (a) Angular spectra of the reflectance R and modulation coefficient $\zeta$ and temporal dependence of $R$ (inset). (b) Spatial distribution of the electromagnetic field $H_y$ (black) and $|\mathbf{E}|^2$ (red) and its modulation $\delta|\mathbf{E}|^2/2<|\mathbf{E}|^2>$ (blue) due to acoustic oscillations inside the structure.

The distribution of electro-magnetic field in Fig.3b demonstrates excitation of a surface mode in the guiding layer of SiC. Exponential decrease of the magnetic field in the air gap and the substrate may be seen. The higher value of the electric field modulation (compare with Fig. 2b) in the prism with respect to modulation of the field in the SiC layer proves the validity of application of the NPS structures.

## Acousto-optical modulation in positive permittivity semiconductor structures

In order to achieve high sensitivity for photo-elastic variation of permittivity we propose using PPS waveguide structures due to lower attenuation and greater Q-factor. Similar to the analyzed structures with SiC, longitudinal acoustic wave is used to modulate both dielectric permittivity of the prism and coupling strength by moving the waveguide cladding (air). However, oppositely to the surface phonon-polaritons, the case of waveguide modes allows to utilize both s- and p-polarized light.

Implementation of waveguides with high contrast of the refraction index, such as air-GaAs-LiF allows achieving deep resonances up to 0.05% reflection and angular width of 0.5° in the prism. In the case of p-polarized light the optimal waveguide configuration is 1.59 µm air gap, 1.8 µm GaAs core ($n$=3.27) and LiF substrate ($n$=1.055) with resonance angle 36.23°. The tolerance of operating wavelength variation is about 4.5 %/nm and sensitivity to variation of GaAs waveguide layer thickness is slightly less than 20 %/nm. In Fig.4a angular distributions of the reflectance and the modulation $\zeta$ are shown, while Fig.4b shows the field distribution inside the structure and its variation. It is seen that application of the PPS structure provides the highest possible value of modulation $\zeta$=100%.

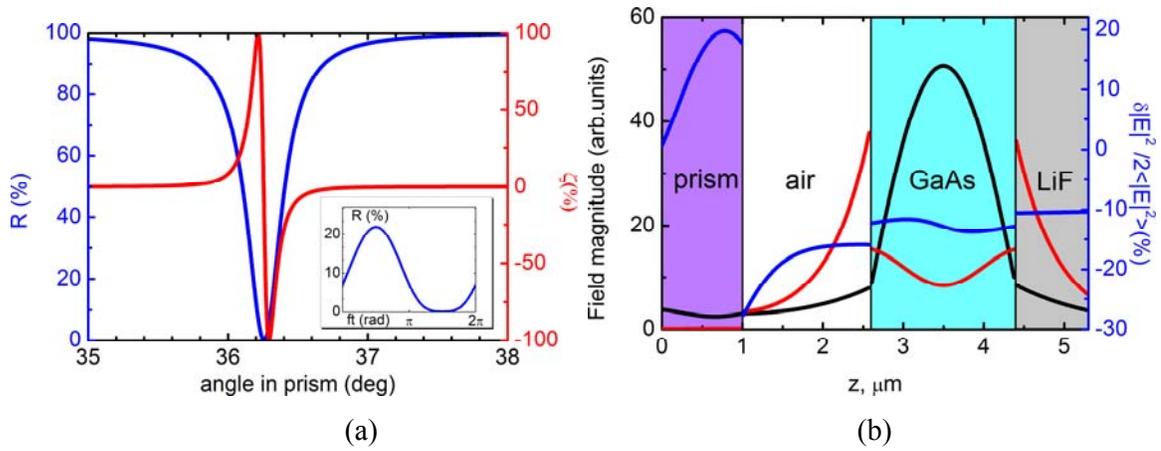

(a)  (b)

Fig.4 (a) Acousto-optical modulation enhanced via excitation of waveguide mode in air-GaAs-LiF waveguide in p-polarized light. (a) Angular spectra of the reflectance $R$ and modulation coefficient $\zeta$ and temporal dependence of $R$ (inset). (b) Spatial distribution of the electromagnetic field $H_y$ (black) and $|\mathbf{E}|^2$ (red) and its modulation $\delta|\mathbf{E}|^2/2\langle|\mathbf{E}|^2\rangle$ (blue) due to acoustic oscillations inside the structure.

Modulation of the reflectance is nonlinear (inset of Fig.4a). The modulated signal acquires square shape rather than sinusoidal one with further increase of the acoustic power density $W$. Due to the higher Fresnel coefficients, s-polarized light provides narrower resonances $\Delta\theta$=0.1°, actually narrower than the value of the shift, caused by the acoustic wave. To achieve 0.05% reflectance minimum at 56.78° we use the PPS structure with 2.33 µm air gap and 1.8 µm GaAs core. Sensitivity to both variations of GaAs waveguide layer thickness and operating wavelength variations is about 4 %/nm. The parameters of this structure are shown in Fig. 5. The modulation of light is nonlinear (see the inset in Fig.5a) and its time dependence also has the square-shape whose duty cycle is controlled by the acoustic power density. In order to obtain linear regime it is necessary to decrease power density by about 6 times.

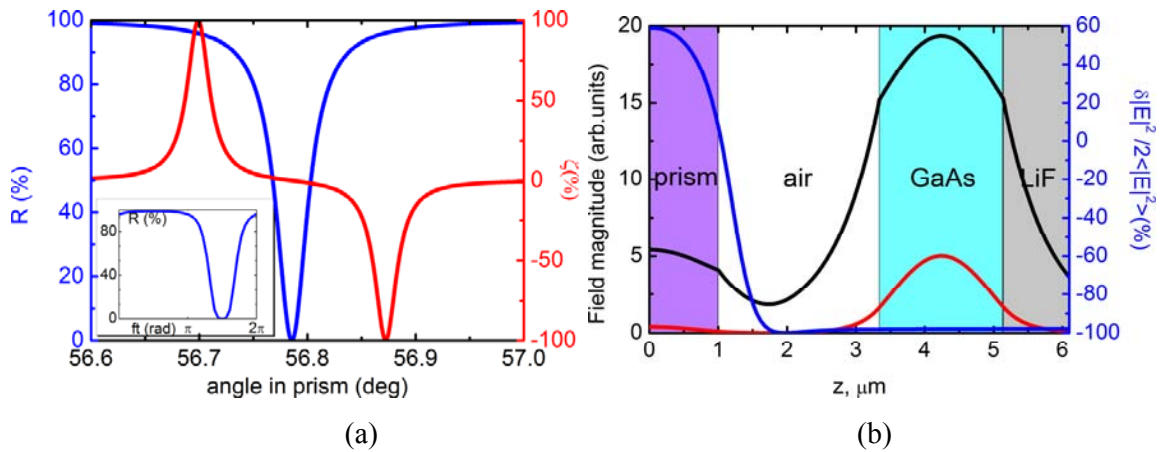

(a)  (b)

Fig.5 Acousto-optical modulation enhanced via excitation of waveguide mode in the air-GaAs-LiF waveguide in s-polarized light. (a) Angular spectra of the reflectance $R$ and modulation coefficient $\zeta$. (b) Spatial distribution of the electromagnetic field $H_y$ (black) and $|\mathbf{E}|^2$ (red) and its modulation $\delta|\mathbf{E}|^2/2<|\mathbf{E}|^2>$ (blue) due to acoustic oscillations inside the structure.

In Fig. 5b huge modulation of the electric field is demonstrated. If we design the similar structure with output optical channel located in the guiding layer it will provide higher values of modulation. In this case modulation efficiency of light will increase with the width of the acoustic impact region. However, an important advantage of the device proposed here will be lost, namely, the possibility of modulation at frequencies above 1 GHz. With increasing width of the acoustic beam, the electric capacitance of the piezoelectric transducer increases. As a result, the excitation of acoustic waves at high frequencies will be extremely difficult and the modulation depth will be low.

## Conclusion

Application of the NPS structures allows achieving modulation of a beam with wide angular spectrum, while the PPS waveguides provide higher modulation coefficient $\zeta$.

Most perspective for acousto-optical modulation in the mid-IR region are waveguide structures with high refractive index contrast. Excitation of TM or TE guided modes produces narrow and deep resonances, which makes it possible to achieve extremely high modulation values $\zeta$. It is worth noting that the values of angular width are given inside the prism so they are more than 3 times smaller than that in the air. Therefore, collimation of a mid-infrared beam for modulation in such structure is an important though quite resolvable task.

With increase in the frequency of ultrasound, acoustic losses increase. Therefore, it is necessary to reduce the propagation length of the acoustic wave $h$ (see Fig. 1). The easiest way is to proportionally reduce the size of the prism. However, in our study, structures with very narrow resonances were considered, so it is necessary to use collimated optical beams. A beam with a minimum angular spectrum can be achieved by focusing the radiation on the prism-gap interface. After calculations in the approximation of weakly diverging beams ( $l \gg \lambda/\pi n$ ), one can obtain:

$$h = l \cos\theta + \sqrt{\frac{2l\lambda}{\pi n}} \sin\theta, \quad (7)$$

where $l$ is light beam waist length, which corresponds to the half path of optical beam in prism, $\lambda$ is light wavelength, $n$ is refractive index of the prism, $\theta$ is angular position of the resonance in the structure. Eq.(7) has solution with respect to $l$ for any small value of $h$. It means that for any small size of prism it is possible to direct a collimated beam at some angle. The approximation is valid for the length of propagation $h > \lambda$ It should be noted that in practice, such dimensions may not be convenient for optical adjustment. However, for dimensions of about $h = 1$ mm, taking into account the attenuation value of 20db/cm at 1 GHz [22], [23], it can be argued that our device can operate at 1 GHz without significant acoustic losses.

Here we demonstrated that in the case of SiC based structures a significant contribution to modulation is provided by the gap modulation $\delta d$ due to the wide resonance bandwidth, while in the case of GaAs layer based waveguides the variation of dielectric permittivity $\delta \varepsilon$ is much more influential. As follows from Eqs. (S4) and (S6) (see Supplementary), with an increase in the acoustic frequency, the contribution of $\delta d$ decreases, and the influence of $\delta \varepsilon$ is independent of frequency. Therefore, for operation at frequencies above 1 GHz the structures with a narrower resonance are preferred as demonstrated for the GaAs layer based waveguides.

Waveguides with lower dielectric contrast such as air-GaAs-CdTe or air-ZnS-LiF were also analyzed. Although the optimal matching conditions for excitation of waveguide resonance with $R_{min} = 0.05\%$ also could be realized, the acousto-optical modulation of the signal would be several times weaker. This is caused by less efficient energy localization in the waveguide core which makes the structure less sensitive to the coupling modulation.

## Acknowledgments

This work was financially supported by Russian Foundation for Basic Research (RFBR) (project N 18-29-20113).